\documentclass[preprintnumbers]{revtex4}
\UseRawInputEncoding
\usepackage{amssymb}
\usepackage{amsmath}
\usepackage{graphicx}
\usepackage{dcolumn}
\usepackage{bm}
\usepackage{subfigure}
\usepackage{color}

\begin{document}

\title{Topology and phase transition for EPYM AdS black hole in thermal potential}
\author{Yun-Zhi Du, Huai-Fan Li\footnote{the corresponding author}, Yu-Bo Ma, Qiang Gu}
\affiliation{Department of Physics, Shan xi Da tong University, Da tong 037009, China\\
Institute of Theoretical Physics, Shan xi Da tong University, Da tong 037009, China}

\thanks{\emph{e-mail:duyzh13@lzu.edu.cn, huaifan999@163.com, yuboma.phy@gmail.com, gudujianghu23@163.com}}

\begin{abstract}
As we all know the local topological properties of thermodynamical systems can be expressed by the winding numbers as the defects. The topological number that is the sum of all winding numbers can be used to classify the global topological nature of thermodynamical systems. In this paper, we construct a kind of thermal potential and then put the Einstein-power-Yang-Mills AdS black hole in it. Through the analysis of the geometric characteristics of the thermal potential based on the complex analysis we find the topological number is an invariant that is same as shown in the way of the Duan's $\phi$-mapping topological current [Sci. Sin. 9, 1072 (1979)]. Furthermore, we adopt the Kramer's escape rate method to investigate the intensity of the first-order phase transition.
\end{abstract}

\maketitle

\section{Introduction}
Since the 70's of last century, black holes have been discovered to be both strong gravitational and thermodynamic systems, satisfying the four laws of thermodynamics. Subsequently, Hawking and Page \cite{Hawking1983} presented that there exists a phase transition between a pure AdS spacetime and a stable large black hole state, i.e., Hawking-Page phase transition, which was explained to a confinement/deconfinement phase transition in the gauge theory \cite{Witten1998}. That makes black hole be paid more and more attention. Since the lack of pressure in the traditional black hole thermodynamics, the authors \cite{Kastor2009} expanded black hole thermodynamics into the expanded phase space though regarding the negative cosmological constant as pressure. These leads that black hole thermodynamics have attracted lots of attention. Especially the thermodynamical properties of AdS black holes in the expanded phase space had been widely considered \cite{Hendi2017a,Hennigar2017a,Frassin,Kubiznak2012,Cai2013,Ma2017,Banerjee2017,Mann1207,Wei2015,Bhattacharya2017,Zeng2017,Zhang1502,Du2021,Zhang2020}. Recently, in order to make black hole thermodynamics more like the ordinary thermal systems, people proposed the holographic thermodynamics \cite{Visser2022,Ahmed2023} and the restricted phase space thermodynamics \cite{Zhao2022,Gao2022,Gao2022a,Du2023} of AdS black holes to give a holographic interpretation of black hole thermodynamics. Furthermore, through the $\phi$-map topological flow theory \cite{Duan1979} the authors proposed black hole solutions are regarded as defects which are described by winding numbers \cite{Wei2022}. The wingding number of the local stable black hole state is one, and for the local unstable black hole state it is negative one. The topological number is the sum of all wingding numbers to reveal the global topological nature of black hole system. All the black hole systems may be classified by topological numbers. In this work we will apply these results to the topological property of the Einstein-power-Yang-Mills AdS black hole.

On the other hand, the study of the phase transition of a thermodynamic system is usually based on the heat capacity, Gibbs free energy, as well as the Maxwell's equal area law. Recently, the authors \cite{Xu2021,Xu2023} proposed a new way to investigate the phase transition of thermodynamic systems, i.e., though constructing a kind of thermal potential by the temperature of the ensemble, black hole systems are put in such unique thermal potential. We can obtain the corresponding thermodynamic properties of black holes directly by the geometric characteristics of the thermal potential. Black holes will be in such a thermal potential by the driven of the thermal fluctuations. Hence the complex structure of the phase transition for black holes can be obtained by the method of the complex analysis. Furthermore, the detailed description of the phase transition process of the charged AdS black holes were also presented in \cite{Xu2023}. In this manuscript, we will construct a kind of thermal potential for the charged AdS black holes with non-linear source, then use the complex analysis method and the results in Ref. \cite{Wei2022} to uncover the nature of topological number for the different black hole systems. And through the Kramer's escape rate method \cite{Risken1988} we will examine the first-order phase transition process to exhibit the corresponding rate of phase transition.

As well we know, for linear charged AdS black holes at the critical point there exists a scaling symmetry \cite{Johnson2018,Johnson2018a}, the entropy is proportional to the square of charge, the pressure is proportional to the minus square of charge, and the temperature is proportional to the negative power of charge. So for the non-linear charged AdS black holes, whether the similar scaling symmetry still exists? Additional, with the existence of the infinite self-energy of point like charges in Maxwell's theory \cite{Born1934,Anninos2009,Cai2008}, the linear theory will bring in non-linearities \cite{Birula1970} as the field is strong. An interesting non-linear generalization of charged AdS black holes involves a Yang-Mill field exponentially coupled to Einstein gravity, i.e., the Einstein-power-Yang-Mills (EPYM) theory. Several thermodynamical features of the EPYM AdS black hole in extended phase space have recently been studied \cite{Du2021,Zhang2015,Moumni2018}.

Inspired by these, we mainly investigate the relation between topological number and phase transitions of the EPYM AdS black hole. This work is organized as follows. In Sec. \ref{scheme2}, we would like to briefly review the construction of the thermal potential. In Sec. \ref{scheme3}, we investigate the topology of the EPYM AdS black holes with different values of model parameters through the complex analysis, and check out the corresponding results as the system undergoing a first-order phase transition by the comments in Ref. \cite{Wei2022}. Furthermore, based on the Kramer's escape rate method \cite{Risken1988} the rates of first-order phase transition processes for this system are also exhibited in Sec. \ref{scheme4}.  A brief summary is given in Sec. \ref{scheme5}.

\section{Construction of thermal potential}
\label{scheme2}

For a canonical ensemble with temperature $T$, it can be consist of all kinds of states, among them one state can be regarded a real black hole in equilibrium when the ensemble temperature equals the black hole hawking temperature. The real black hole state is one of the solutions for the Einstein field equation while others are not. For a black hole thermodynamic system, we can construct the thermal potential \cite{Xu2021,Xu2023}
\begin{eqnarray}
U=\int \left(T_+-T\right)dS, \label{U}
\end{eqnarray}
where $T_+$ and $S$ are the hawking temperature and the entropy of black hole, respectively. The ensemble temperature $T$ is treated as an independent constant, which can take any positive value in any way. For a black hole, its hawking temperature $T_+$ can be written as $T_+(S,~q,~\Lambda,~J,~...)$. In this work, we just consider the case of $T=T_+$. In addition, according to the first law of thermodynamics $dM=T_+dS-PdV+\phi dq+...$, the thermal potential becomes
\begin{eqnarray}
U=M+PV-TS.
\end{eqnarray}
Formally, we can see that the thermal potential is equivalent to the free energy landscape of AdS black hole systems.

The integrand in the above definition (\ref{U}) of thermal potential can be understood as the deviation of all possible states in the canonical ensemble from the real black hole state in equilibrium. In other words, in the equilibrium state, the thermal potential exhibit the following extreme behavior,
\begin{eqnarray}
\frac{dU}{dS}=0,~~~ \text{or} ~~~T=T_+.
\end{eqnarray}
Through the construction of the thermal potential (\ref{U}), we can put a black hole in the thermal potential field $U$. Due to the thermodynamic fluctuations, the black hole will exhibit the stochastic behavior in this thermal potential field, which can reflect some thermodynamic characteristics of black holes.

\begin{figure}[htp]
\centering
\includegraphics[width=0.32\textwidth]{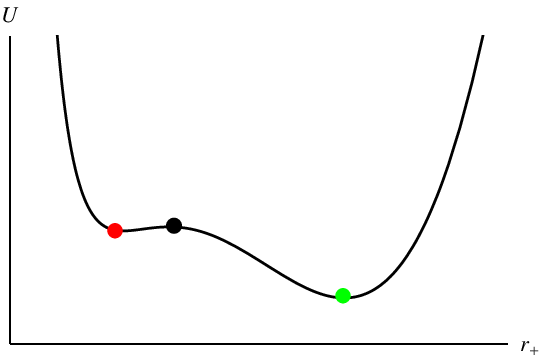}
\caption{The schematic diagram of thermal potential, where the green-point represents the global minimum (the large black hole), the black-point represents the local maximum (the middle black hole) and the red-point represents the local minimum (the small black hole). Except for these three extreme points, the other positions stand for unknown states. }\label{Uexample}
\end{figure}

Moreover, we can investigate the stability of black hole by this thermal potential, in which the extreme minimum stands for the stable state and the local extreme maximum is for the unstable state, i.e.,
\begin{eqnarray}
\delta\left(\frac{dU}{dS}\right)_{T=T_+}=\left(\frac{\partial T_+(S,~q,~J,~\Lambda,...)}{\partial S}\right)_{q,~J,~\Lambda,~...}\delta S.
\end{eqnarray}
When the right hand of the above equation is positive, the black hole will be in stable state, while the negative right hand corresponds to the unstable state, which are shown in Fig. \ref{Uexample}. It is obviously that the lowest point (the green point) is the most stable state in the entire canonical ensemble. For different parameters of the black hole, the extreme point of the thermal potential constantly changes, which corresponds to the changes between the black hole state and other unknown states. In the framework, the authors had investigated the microscopic phase transition mechanism of the charged AdS black holes and found that the phase transition of the large and small black holes exhibits severely asymmetric features, which fills the gap in the analysis of stochastic processes in the first-order phase transition rate problem of AdS black holes \cite{Xu2023}.

\section{Topology of EPYM AdS black hole system}
\label{scheme3}
In the four-dimensional AdS spacetime with the Einstein-power-Yang-Mills gravity, the action had been given in Refs. \cite{Zhang2015,Corda2011,Mazharimousavi2009,Lorenci2002}
\begin{eqnarray}
I=\frac{1}{2}\int d^4x\sqrt{g}\left(R-2\Lambda-\mathcal{F}^\gamma\right)
\end{eqnarray}
with the Yang-Mills(YM) invariant $\mathcal{F}$ and the YM field $F^{(a)}_{\mu\nu}$
\begin{eqnarray}
\mathcal{F}&=&Tr\left[F^{(a)}_{\mu\nu}F^{(a)\mu\nu}\right],\\
F^{(a)}_{\mu\nu}&=&\partial_\mu A^{(a)}_\nu-\partial_\nu A^{(a)}_\mu+\frac{1}{2\xi}C^{(a)}_{(b)(c)}A^{(b)}_\mu A^{(c)}_\nu.
\end{eqnarray}
Here, $Tr\left[F^{(a)}_{\mu\nu}F^{(a)\mu\nu}\right]=\sum^3_{\mu\nu}F^{(a)}_{\mu\nu}F^{(a)\mu\nu}$,
$R$ and $\gamma$ are the scalar curvature and a positive real parameter, respectively, $C^{(a)}_{(b)(c)}$ represent the structure constants of three-parameter Lie group $G$, $\xi$ is the coupling constant, and $A^{(a)}_\mu$ represents the SO(3) gauge group Yang-Mills (YM) potentials defining by the Wu-Yang (WY) ansatz \cite{Balakin2016,Balakin2007,Balaki2007}. Variation of the action with respect to the spacetime metric $g_{\mu\nu}$ yields the field equations
\begin{eqnarray}
&&G^{\mu}_{~\nu}+\Lambda\delta^{\mu}_{~\nu}=T^{\mu}_{~\nu},\\
&&T^{\mu}_{~\nu}=-\frac{1}{2}\left(\delta^{\mu}_{~\nu}\mathcal{F}^\gamma-4\gamma Tr\left[F^{(a)}_{\nu\rho}F^{(a)\mu\rho}\right]\mathcal{F}^{\gamma-1}\right).
\end{eqnarray}
It is obviously that the EPYM theory with $\gamma=1$ can reduce to the standard Einstein-Yang-Mills theory \cite{Habib2007,Mazharimousavi2008}. In the following we mainly pay attention on the role of the non-linear YM charge parameter $\gamma$.
With the WY ansatz for the YM field, the invariant for this field becomes
\begin{eqnarray}
Tr\left[F^{(a)}_{\mu\rho}F^{(a)\mu\rho}\right]=\frac{q^2}{r^4},
\end{eqnarray}
which will lead to the disappearance of the structure constants described the non-Abelian property of the YM gauge field.

The four-dimensional static spherical symmetry EPYM black hole solution in the AdS spacetime reads \cite{Yerra2019}
\begin{eqnarray}
f(r)=1-\frac{2M}{r}-\frac{\Lambda}{3}r^2+\frac{(2q^2)^\gamma}{2(4\gamma-3)r^{4\gamma-2}}.
\end{eqnarray}
Here, $q$ is the YM charge, the non-linear YM charge parameter satisfies $\gamma\neq0.75$ and $\gamma>0$. In the extended phase space the cosmological constant $\Lambda$ is interpreted as the pressure $P=-\frac{\Lambda}{8\pi}$, whose dual quantity is volume $V=\frac{4\pi r_+^3}{3}$. Since the YM charge term of this theory with $\gamma=1$ have the same form as the Maxwell charge term in the Einstein-Maxwell-Yang-Mills (EMYM) theory, the role of the YM charge will be similar as that of the Maxwell charge in EMYM theory. The only difference between them is that they have different gauge groups, i.e., the YM field is of SO(3), the Maxwell field is of U(1). These results are consistent with that in Ref. \cite{Moumni2018}. With the above metric function, the black hole hawking temperature, mass parameter, and entropy are, respectively,
\begin{eqnarray}
T_+&=&\frac{1}{4\pi r_+}\left(1+8\pi P r_+^2-\frac{(2q^2)^\gamma}{2r_+^{4\gamma-2}}\right),~S=\pi r_+^2\\
M&=&\frac{r_+}{2}\left(1+\frac{8\pi P r_+^2}{3}+\frac{(2q^2)^\gamma}{2(4\gamma-3)r_+^{4\gamma-2}}\right).
\end{eqnarray}
Therefore, the thermal potential of the EPYM AdS black hole is expressed as
\begin{eqnarray}
U&=&\int \left(T_+-T\right)dS\nonumber\\
&=&\frac{1}{2}r_+-T\pi r_+^2+\frac{4}{3}P \pi r_+^3-\frac{ \left(2q^2\right)^{\gamma }}{4(3-4 \gamma)}r_+^{3-4 \gamma }
\end{eqnarray}
Since the various black hole states are located at the extreme points of the thermal potential function, so we have
\begin{eqnarray}
h(r_+)=\frac{dU(r_+)}{dS(r_+)}=-T+\frac{1}{4 \pi r_+}+2 P r_+-\frac{(2q^2)^{\gamma }}{8\pi  }r_+^{1-4 \gamma }
\end{eqnarray}
We can transform the black hole thermodynamic properties into solving the zero-points of the above function $h(r_+)$. To investigate the whole zero-points properties of the thermal potential function, the method of complex analysis should be applied to make the real thermal potential to the one with complex continuation, i.e., $h(r_+)\rightarrow h(z)$. The different black hole states correspond to the positive real zero-points of $h(z)$. Its analytical function with complex continuation has the following form
\begin{eqnarray}
h(z)=\frac{1}{z^{4\gamma-1}}\left(-Tz^{4\gamma-1}+\frac{z^{4\gamma-2}}{4\pi}+2Pz^{4\gamma}-\frac{(2q^2)^\gamma}{8\pi}\right).\label{h}
\end{eqnarray}

In the complex analysis, the Argument Principle is often applied to calculate the number of zero-points for analytic functions. For a meromorphic function $h(z)$ in a closed cuntour $C$, which is analytically nonzero on $C$, we have
\begin{eqnarray}
N(h,C)-P(h,C)=\frac{1}{2\pi i} \oint_C\frac{h'(z)}{h(z)}dz=\frac{\Delta_C arg~h(z)}{2\pi},
\end{eqnarray}
where $N(h,C)$ and $P(h,C)$ are the number of zero-points and poles for the analytic function $h(z)$, $arg~h(z)$ stands for the argument of $h(z)$. With a transformation $\omega=h(z)$, the above equation can be rewritten as the number of rotations of $\omega$ around the origin of $C'$ along with the complex variable $z$ moving around $C$. Here $C'$ is the image curve of $C$ after this transformation. Thus the topological number can be denoted as
\begin{eqnarray}
W:=\frac{1}{2\pi i}\oint_{C'}\frac{d\omega}{\omega}=\frac{1}{2\pi i}\oint_{C'}\frac{h'(z)}{h(z)}dz.
\end{eqnarray}
Specially, for the analytic function $h(z)$ without poles inside $C$, i.e., $P(h,C)=0$, the topological number equals to the number of zero-points $W=N(h,C)$. Here we should point out that $N(h,C)$ is determined by the order of the complex various $z$. For the AdS EPYM black hole system, for convenience we require that the powers of all order of $z$ must be positive integers, i.e., $\gamma=\frac{n+2}{4}$ with the positive integer $n$ ($n>1$). The equation (\ref{h}) becomes
\begin{eqnarray}
h(z)=\frac{1}{z^{n+1}}\left(-Tz^{n+1}+\frac{z^{n}}{4\pi}+2Pz^{n+2}-\frac{(2q^2)^{n+2}}{8\pi}\right),
\label{h}
\end{eqnarray}
which have $n+2$ zero-points and a $n+1$-order pole in the entire complex plane. Thus the topological number of this system is $W=1$ for different values of $n$, i.e., it is independent with the non-linear YM charge parameter. Note that for some certain values of the pressure and YM charge, the EPYM AdS black hole will reduce to different black hole systems and the corresponding topological number will change for different black hole systems (see as Tab. \ref{tcp}). These results are consistent with that in Refs. \cite{Wei2022,Fang2023}
\begin{table}[htbp]
\centering
\caption{Topological numbers of different black hole system with different values of $P,~q$}
\begin{tabular}{c|c|c|c|c}
\hline\hline
\centering
$P,~q$~~&$P=q=0$ Sch-BH~~& $P=0,~q\neq0$ RN-BH~~&$P\neq0,~q=0$ Sch-AdS-BH~~&$P\neq0,~q\neq0$ EPYM-AdS-BH~~\\ \hline
~~$N(h,C)$ ~~&0~~&$n+1$~~&2~~&$n+2$~~\\ \hline
$P(h,C)$~~&1~~&$n+1$~~&1~~&$n+1$~~\\ \hline
$W$~~&-1~~& 0~~& 1~~& 1~~\\ \hline \hline
\end{tabular}\label{tcp}
\end{table}

In the following, we will verify the above viewpoint. For AdS EPYM black holes, for any values of the non-linear YM charge parameter the critical points read as \cite{Du2021}
\begin{eqnarray}
(r_+^c)^{4\gamma-2}&=&(2q^2)^\gamma f(1,\gamma),~~~f(1,\gamma)=\gamma(4\gamma-1),\\
T_+^c&=&\frac{2\gamma-1}{4\gamma-1}\frac{1}{\pi(2q^2)^{\gamma/4\gamma-2}f^{1/4\gamma-2}(1,\gamma)},\\
P^c&=&\frac{2\gamma-1}{16\pi\gamma(2q^2)^{\gamma/2\gamma-1}f^{1/2\gamma-1}(1,\gamma)},\\
G^c&=&M(r_+^c,~P^c)-T_+^cS^c.
\end{eqnarray}
For the sake of discussion, we introduce the dimensionless thermodynamic quantities as follows
\begin{eqnarray}
p\equiv\frac{P}{P^c},~~~~t\equiv\frac{T}{T_+^c},~~~~x\equiv\frac{r_+}{r_+^c},~~~~
s\equiv\frac{S}{S^c},~~~~u\equiv\frac{U}{U^c},~~~~g\equiv\frac{G}{G^c}.
\end{eqnarray}
Here $U^c=\frac{1}{CQ^\gamma}\left(-\frac{12(2\gamma-1)}{4\gamma-1}+6Q^{2\gamma(1-\gamma)}
+\frac{2\gamma-1}{\gamma}Q^{6\gamma(1-\gamma)}+\frac{3}{4\gamma-3}C^{4(\gamma-1)})\right)$ with $C=\left[\gamma(4\gamma-1)\right]^{1/2(1-\gamma)}$ and $Q=(2q^2)^{\frac{1}{2(1-2\gamma)}}$. Thus the thermal potential reads
\begin{eqnarray}
u=\frac{1}{CQ^\gamma U^c}\left(-\frac{12(2\gamma-1)}{4\gamma-1}tx^2+6Q^{2\gamma(1-\gamma)}x
+\frac{2\gamma-1}{\gamma}Q^{6\gamma(1-\gamma)}px^3+\frac{3}{4\gamma-3}C^{4(\gamma-1)})x^{3-4\gamma}\right).\label{ur}
\end{eqnarray}
From Eq. (\ref{ur}) we can seen that there exist two key parameters ($p$ and $t$) affecting the behaviors of the thermal potential.
Generally, there are two methods to obtain the information of the phase transition: one is the Gibbs free energy, another is the Maxwell's equal area law. Here we adopt the first method to investigate the property of phase transition for this system. On the other hand the authors \cite{Wei2022,Fang2023} proposed that the local topological properties of thermodynamical systems can be reflected by the winding numbers as the defects, i.e., for the local stable black hole state the winding number equals to one, for the local unstable black hole state it becomes minus one. The global topological nature can be classified by the topological number which is the sum of all local winding numbers. In the following we will apply indirectly these results to this system.

\begin{figure}[htp]
\subfigure[$p=0.125$]{\includegraphics[width=0.32\textwidth]{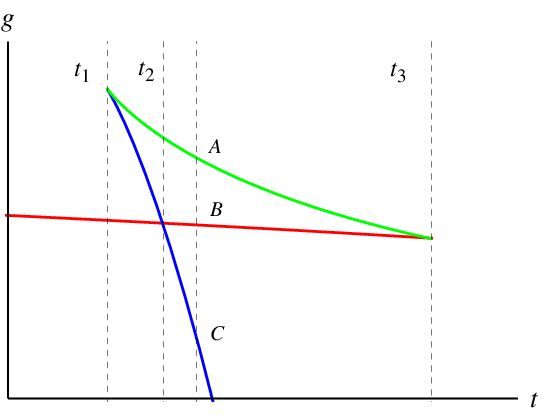}\label{gt}}~~
\subfigure[$p=0.125,~t=t_1$]{\includegraphics[width=0.32\textwidth]{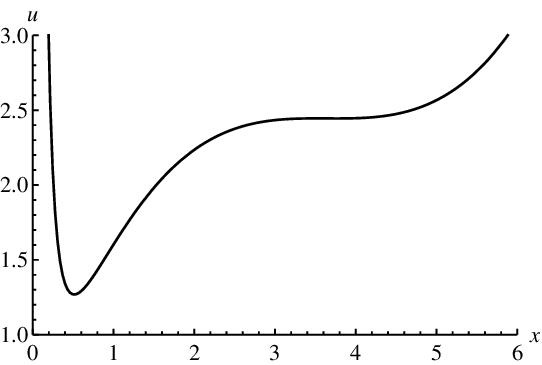}\label{ux3}}
\subfigure[$p=0.125,~t_1<t<t_2$]{\includegraphics[width=0.32\textwidth]{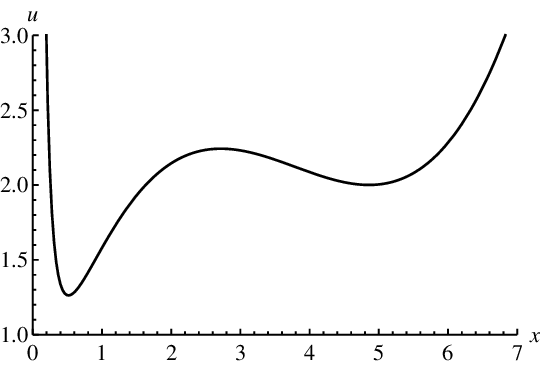}\label{ux2}}\\
\subfigure[$p=0.125,~t=t_2$]{\includegraphics[width=0.32\textwidth]{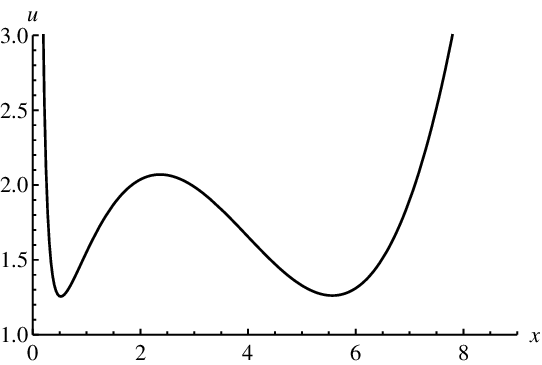}\label{ux}}~~
\subfigure[$p=0.125,~t_2<t<t_3$]{\includegraphics[width=0.32\textwidth]{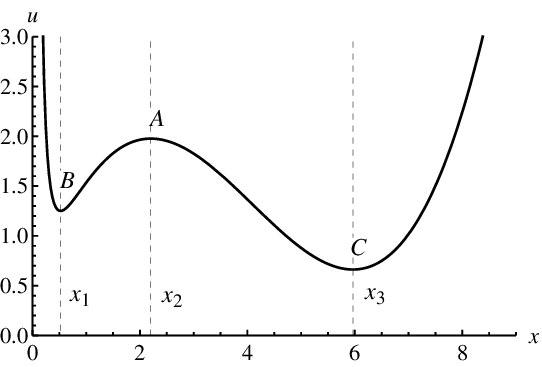}\label{ux1}}~~
\subfigure[$p=0.125,~t=t_3$]{\includegraphics[width=0.32\textwidth]{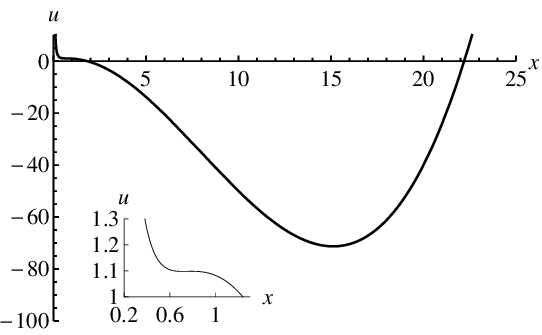}\label{ux4}}
\caption{The plots of the Gibbs free energy and the thermal potential versus temperature with different values of the dimensionless pressure. In $g-t$ plane, the red line is the small black hole branch, the green line is the middle black hole branch, and the blue line is the large black hole branch. The parameters are set to $\gamma=1.25,~q=1/\sqrt{2}$.  }\label{gtux}
\end{figure}

The swallowtail behavior in the $g-t$ phase diagram and the corresponding black hole states with different temperatures in $u-x$ planes are exhibited in Fig. \ref{gtux}. As $t=t_1$ shown in Fig. \ref{gt} the free energies of the large and middle black hole states are the same, for the small black hole state it is lowest that indicates the system is in the stable small black hole state, while as $t=t_3$ the free energies of the small and middle black hole states are the same, the large black hole state is lowest. The corresponding results are expressed in Figs. \ref{ux3} and \ref{ux4}: the thermal potential only has a global minimum that stands for the stable small black hole state when $t=t_1$ or that stands for the stable large black hole state as $t=t_3$. For these two cases there is one winding number of the stable black hole state, so the topological number equals to one. From Fig. \ref{ux} it is obviously that when $t=t_2$ (i.e., the first-order phase transition point) the thermal potential have two global minima and a local maximum, and the thermal potentials of two minima equal to each other. That is to say the system is in the coexistence of the stable large and small black hole states, while the middle black hole state is local unstable. Hence there are three winging numbers: 1, 1, -1, and he topological number of this system undergoing a first-order phase transition is one ($1+1-1=1$). For other two cases ($t_1<t<t_2$ and $t_2<t<t_3$) shown in Figs. \ref{ux2} and \ref{ux1}, the thermal potential have two local stable minima and one local unstable maximum, the corresponding winding numbers of these states are 1, 1, -1 and the topological number is one. Though above analysis, for a given pressure no matter whether the system undergos a phase transition, the topological number is an invariant ($W=1$). The results are the same as that obtained by the complex analysis.

\section{Rate of the first-order phase transition}
\label{scheme4}
\begin{figure}[htp]
\subfigure[$p=0.125$]{\includegraphics[width=0.32\textwidth]{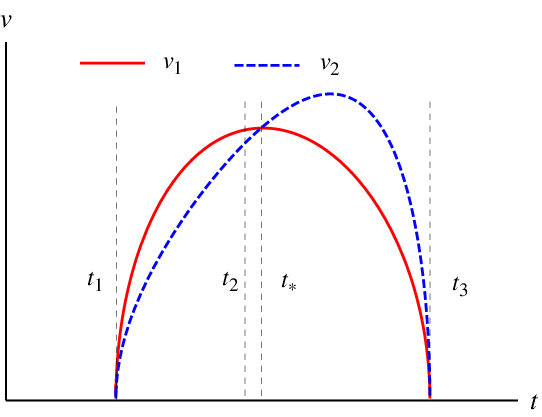}\label{v1}}~~~~~~
\subfigure[$p=0.125$]{\includegraphics[width=0.32\textwidth]{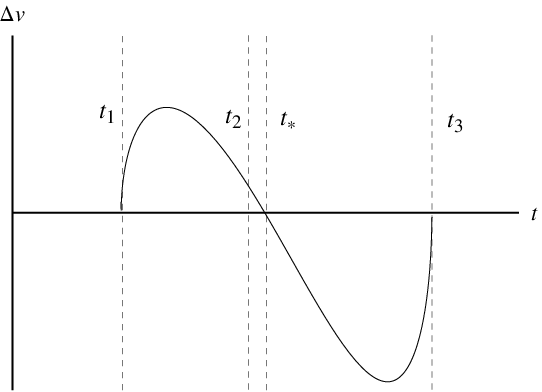}\label{dv1}}
\caption{The transition rates $\nu_{1,2}$ and their difference $\Delta\nu$ with respect to the ensemble temperature $t$ at the pressure $p=0.125$. The parameters are set to $\gamma=1.25$, $q=1/\sqrt{2}$, and $D=20$. }\label{v}
\end{figure}

Based on the black hole molecular hypothesis \cite{Wei2015}, the black hole phase transition is the rearrangement of black hole molecules due to the thermal fluctuation in the thermal potential. Fig. \ref{ux1} might describe a molecular rearrangement. We assume that the ensemble temperature is much lower than the barrier height. Molecules will spend a lot of time near the potential minimum (B), and only rarely will Brownian motion take them to the top of the barrier (A). Once the molecule reaches the top of the barrier, it is likely to fall equally to either side of the barrier. If it moves to the right-hand side, it will rapidly fall to the other minimum (C), stay there for a while, and then perhaps cross back to the original minimum (B). In order to depict the crossing rate (B$\rightarrow$A or C$\rightarrow$A), we introduce the Kramer's rate \cite{Risken1988}
\begin{eqnarray}
\nu=\frac{\sqrt{|u''(x_{min})u''(x_{max})|}}{2\pi}e^{-\frac{x_{max}-x_{min}}{D}},
\end{eqnarray}
where $D$ is the constant diffusion coefficient, $x_{max}$ and $x_{min}$ are the corresponding locations of the thermal potential extreme points and they can be obtained by solving the equation of $u(x)'=0$. $''$ stands for the second derivative of the thermal potential $u(x)$. We define the crossing rate from B to A (or from C to A) as $\nu_1$ (or $\nu_2$)
\begin{eqnarray}
\nu_1=\frac{\sqrt{|u''(x_1)u''(x_2)|}}{2\pi}e^{-\frac{x_2-x_1}{D}},~~
\nu_2=\frac{\sqrt{|u''(x_3)u''(x_2)|}}{2\pi}e^{-\frac{x_3-x_2}{D}}.
\end{eqnarray}
Namely, $\nu_1$ denotes the rate of phase transition from the small black hole to the large black hole, $\nu_2$ is the one from the large black hole to the small black hole. The phase transition rates and their difference ($\Delta\nu=\nu_1-\nu_2$) for different black hole states are depicted in Fig. \ref{v}. When the ensemble temperature equals $t_1$ or $t_3$, the phase transition rate is zero, which means that no phase transition emerges. With the increasing of the ensemble temperature from $t_1$ to $t_3$, both two rates show a trend of increasing first and then decreasing. At $t=t_2$ the depths of thermal potential between two minima and local maximum, while the corresponding phase transition rates are not equal to each other ($\nu_1>\nu_2$).  This indicates the transition process from the small black hole to the large black hole is far from that of the transition from the large black hole to the small black hole. Two rates will be the same as $t=t_*$, which indicates the phase transition reaches a dynamic equilibrium at this time. The corresponding difference of two rates is positive for $t_1<t<t_*$, it is zero as $t_t*$ and is negative in the range of $t_*<t<t_3$. In short, the phase transition between different black holes presents very asymmetric features, and the overall process of this system initial being in the stable large black hole state is dominated by the phase transition from a small black hole to a large black hole. On the contrary, if the system is initially in the stable small black hole state the phase transition becomes the one from the large black hole state to the small black hole state.

\section{Discussions and Conclusions}
\label{scheme5}
Different black hole solutions are the different defects from the viewpoint of thermodynamics, which can be characterized by different topological numbers and belong to different topological classes. In this manuscript, we considered the EPYM AdS black hole in a thermal bar with the thermal potential and the whole system was in thermodynamics equilibrium. Based on the construction of the thermal potential, we probed the topology of the whole system with different black hole states by the method of Argument Principle. When the system with $q=0,~\Lambda=0$ reduces to the Schwarzschild black hole, the topological number is negative one. For the RN black hole system (i.e, $q=0,~\Lambda<0$) the topological number is zero. For the charged AdS black hole systems (i.e, $q\neq0,~\Lambda<0$), for example RN-AdS black hole and EPYM AdS black hole, the topological number is one.

Next we checked out these topological numbers by the method of the Duan's $\phi$-mapping topological current when the system was in different states. In other words, for a certain black hole system its topological number is an invariant and is independent of whether the system undergoes a phase transition. Finally, in order to depict the direction of phase transition we introduced the Kramer's rate to investigate the crossing rate between local stable large/small black hole states and local unstable middle black hole state. The results indicated that if the system is initial in the stable large black hole state, the process of phase transition is the from the small black hole state to the large black hole state, and vice versa.

\section*{Acknowledgements}

We would like to thank Profs. Meng-Sen Ma and Zen-Ming Xu for their indispensable discussions and comments. This work was supported by the National Natural Science Foundation of China (Grant No. 12075143, 12375050), the Natural Science Foundation of Shanxi Province, China (Grant No. 202203021221209), and the Teaching Reform Project of Shanxi Datong Universtiy (Grant No. XJG2022234).

\end{document}